\documentclass[%
 reprint,
 amsmath,amssymb,
 aps,
prb,
]{revtex4-2}

\usepackage{float}
\usepackage{braket}
\usepackage{graphicx}
\usepackage{dcolumn}
\usepackage{bm}
\usepackage{inputenc}
\usepackage{blindtext}
\usepackage{color}
\usepackage{booktabs,siunitx}
\usepackage{comment}
\usepackage{placeins}
\usepackage{afterpage}
\usepackage{hyperref}
\hypersetup{colorlinks,citecolor=blue}
\usepackage{todonotes}

\usepackage{tabularx}

\newcolumntype{~}{>{\global\let\currentrowstyle\relax}}
\newcolumntype{^}{>{\currentrowstyle}}

\DeclareMathAlphabet\mathbfcal{OMS}{cmsy}{b}{n}

\renewcommand\vec{\mathbf}

\begin{document}

\title{Effective Gilbert damping in the stochastic Landau-Lifshitz-Gilbert equation}
\author{Mexx E.Y. Regout}
\affiliation{Nanomat and TOM groups, Q-MAT research center and European Theoretical Spectroscopy Facility, Université de Liège, allée du 6 août, 19, B-4000 Liège, Belgium}
\author{Bertrand Dupé}
\affiliation{TOM group, Q-MAT research center, Université de Liège, allée du 6 août, 19, B-4000 Liège, Belgium}
\author{Matthieu J. Verstraete}
\affiliation{Nanomat group, Q-MAT research center and European Theoretical Spectroscopy Facility, Université de Liège, allée du 6 août, 19, B-4000 Liège, Belgium}
\affiliation{ITP, Physics Department, Utrecht University 3508 TA Utrecht, The Netherlands}

\date{May 8th 2026}

\begin{abstract}

Quasi particle based (e.g. Boltzmann equation) studies of spin wave transport often assume that their scattering rates follow the simple form $\eta=\alpha \omega$, with the Gilbert damping $\alpha$ and frequency $\omega$. 
In this work, we examine the effective damping $\alpha_{eff,T}=\eta/\omega$ observed in atomistic spin dynamics, when temperature and spin wave interactions are introduced for a 1D spin chain. We extract the dynamical correlation functions from spin trajectories propagated using the stochastic Landau-Lifshitz-Gilbert equation, and fit the dynamical structure factor, yielding the dispersion and scattering rates for a wide range of temperatures. The resulting effective damping can be very different from the initially constant Gilbert value. It exhibits a temperature and crystal momentum scaling which we explain based on interactions with the Gilbert bath and spin wave scattering by changes in local magnetic order. 
\end{abstract}

\maketitle

\section{Introduction}

Magnons or Spin Waves (SW) are central to the new field of magnonics \cite{Chumak2015-ki}, which aims at using spin currents to transport and process information without the drawbacks of electrical current dissipation. SW can also be used directly to carry out logic operations, as they can show constructive or destructive interference. In these applications, the lifetime of SW is of crucial importance.
As collective spin excitations, the properties of SW are intimately linked to the underlying magnetic order, interactions within a material and dissipation mechanisms.

Magnetic orders and interactions are partially reflected in the SW dispersion relations, which can be obtained from both experiment and theory. On the one hand, the SW dispersions can be measured using inelastic neutron scattering, Raman or Brillouin light scattering (BLS) \cite{Serga2010-mh}. BLS is a versatile method to produce and detect SW in the GHz regime \cite{Borovik-Romanov1982-bu} which makes it ideal in ferromagnets (FM).
On the other hand, SW dispersions can be calculated using time-dependent matrix product states \cite{TDMPS-LSWT-ASD}, linear spin wave theory \cite{fundamentals_of_magnonics} or atomistic spin dynamics (ASD) \cite{Gilbert2004-yp}.

ASD is a versatile real space real time method that has gained interest thanks to the field of skyrmionics, which studies topological non-collinear magnetic textures of nanometric size, at equilibrium \cite{Romming2013-oh,Dupe2014-oh} and out of equilibrium \cite{Buttner2021-jm}. Furthermore, ASD is a versatile method to study magnetization dynamics under external stimuli such as temperature \cite{Ritzmann2014-lp,Evans2012-jt}, magnetic field pulses \cite{Ritzmann2015-xe} and torques of various symmetries \cite{Ritzmann2018-ch}. In the context of SW, ASD has the advantage of enabling the study of short wave length excitations. Finally, this method relies on the parametrization of an extended Heisenberg Hamiltonian, which can be achieved through both neutron scattering experiments \cite{Fishman2018-na} or Density Function Theory calculations \cite{Hoffmann2020-zl, LKAG}.

Based on this Hamiltonian, ASD produces spin trajectories by propagating the Landau-Lifshitz-Gilbert equation (LLG) \cite{Gilbert2004-yp}. The LLG equation approximates the quantum mechanical problem of spin propagation classically. In the context of this work, using ASD with the LLG equation offers the possibility to explore the SW lifetime as a function of both external stimuli (temperature and magnetic field) and the SW wave-vector.

The details of the SW damping has received attention recently in the context of laser-induced dynamics, ferromagnetic resonance and spin caloritronics \cite{Azzawi2017-nx}. The damping has been measured as a function of temperature \cite{Temperature_dependence_effective_gilbert_damping_FeRh} and laser-induced demagnetization \cite{Laser_induced_effective_damping_Thin_film}, and can vary significantly. Nevertheless, in the LLG equation, a single damping parameter $\alpha$ summarizes the dissipation of energy via different channels \cite{Suhl1998-tk}. Although this parameter is usually obtained phenomenologically, it can be computed based on a tight-binding Hamiltonian \cite{Starikov2010-oe,Freimuth2017-oi}, Time Dependent Density Functional Theory (TDDFT) \cite{Capelle2003,Buczek2011} or Korringa Kohn Rostocker \cite{Ebert2011,Lounis2011} methods. Since this parameter can vary by several orders of magnitude, $\alpha$ is essential to quantify SW lifetimes and understand transport of magnetization and heat.

In order to obtain the thermal transport (conductivity $\kappa_{sw}$) for spin waves, two popular approaches are the Green-Kubo equilibrium ASD and the quasi-particle based Boltzmann transport equation (BTE). Both methods have their own advantages in describing the heat transport. The Green-Kubo approach encompasses all the spin-related energy fluctuations in a single value of $\kappa_{sw}$ and has been successfully applied for the prototypical BCC Iron \cite{GKEASD}. One drawback of this approach is the lack of intuition, hidden by a single global value of $\kappa_{sw}$. The integrated autocorrelation function can highlight short and long time dynamics but not how these timescales arise in detail. On the other hand, the BTE is based on a quasi-particle approximation, and often uses a single relaxation time for each state. In compensation, it provides more physical interpretation for the value of $\kappa_{sw}$ as a sum over individual quasi particles and transitions, combining their scattering rate ($\eta_{q}$) and group velocity ($v_{g}$). 
Empirically, the BTE tends to overestimate $\kappa_{sw}$ due to the QP approximation and an overestimation of the group velocity\cite{GKEASD,Magnon_phonon_spin_lattice_Wu}. \\
\\
In this work, we investigate numerically the scattering rate and dispersion of spin waves across a range of wave vectors, with particular attention to their temperature dependence. By employing atomistic spin dynamics simulations (stochastic Landau-Lifshitz-Gilbert equation), we aim to disentangle the dominant relaxation pathways and characterize the scattering rates, dispersions and spectral shapes of the spin wave modes in a 1D chain. Our findings provide insights into how the Gilbert damping, thermostat and spin wave interactions contribute to the scattering rate and dispersions.

\section{Theory}

\subsection{Spin wave dispersion and scattering rate in the linearized LLG}
In order to describe the SW dispersion and scattering rate we first need to define the classical Heisenberg Hamiltonian 
\begin{equation}
\label{spinhamiltonian}
    \mathcal{H} = -\sum_{i,j} J_{ij} \vec{m}_{i} \cdot \vec{m}_{j} -\mu_{m}\sum_{i} \vec{B}\cdot\vec{m}_{i} ,
\end{equation}
where $J_{ij}$ is the Heisenberg exchange between sites $i$ and $j$, which is in the atomic basis (i.e. $J_{ij}$ and $J_{ji}$ are both counted). In our case we work with the nearest-neighbor interaction $J_{1}$. $\Vec{B}=(0,0,B_{z})$ is the external magnetic field (in Tesla) and $\mu_{m}$ is the magnetic moment of the spin (in Bohr magneton). The spin vector $\vec{m}_{i} = (m_{i}^{x},m_{i}^{y},m_{i}^{z})$ has norm  $|\vec{m}_{i}|=1$. In order to model transverse damping we employ the phenomenological Landau-Lifshitz-Gilbert (LLG) equation \cite{Gilbert2004-yp} which is defined as
\begin{equation}
\label{llg}
    \frac{d\vec{m}_{i}}{dt} = -\frac{1}{(1+\alpha^{2})\hbar}(\vec{m}_{i} \times (\vec{H}^{eff}_{i} + \alpha(\vec{m}_{i} \times\vec{H}^{eff}_{i}))),
\end{equation}
with the effective field $\vec{H}^{eff}_{i} = -\frac{\partial \mathcal{H}}{\partial\vec{m}_{i}}$, Gilbert damping $\alpha$ and reduced Planck constant $\hbar$. When considering small fluctuations around the quantization axis (in this case the +z direction), we can linearize the LLG equation assuming $m^{z}_{i}\approx 1$. In the linearized LLG regime we can obtain the analytical solution for the evolution of the spins which results (in one dimension) in the spin wave dispersion 

\begin{align}
\label{dispersion_T0}
    \hbar\omega(q) = \frac{4J_{1}(1-\cos(qa)) + \mu_{m}B_{z}}{1+\alpha^{2}}, 
\end{align}
and scattering rate (which is the inverse lifetime)
\begin{align}
\label{scattering_rate_T0}
    \eta({q}) = \alpha \omega({q})=\tau({q})^{-1}.
\end{align}

The spin wave scattering rate in Eq.~(\ref{scattering_rate_T0}) is a common approximation when interacting with a thermostat (representing phonons, electrons, etc...) \vspace{0.1mm}\cite{FeRh_lifetimes,magnon_transport_ulrike,Magnon-phonon-YIG-gilbert-scattering,Magnon-spin-transport-YIG-insulator}. The validity of this scattering rate form, in realistic models with temperature, is not guaranteed since the spin wave and bath interactions will change\cite{LLB_lifetime,LLB_Power_spectrum}. Besides the temperature component, an extension to include non-local effects in the Gilbert damping $\alpha(q)$ has been made which has shown significant enhancement of the scattering rate in materials like Nickel and Cobalt
\cite{Non_local_damping_for_magnons}. 

\subsection{Dynamical structure factor}

The correlations between spin fluctuations, in space and time, are computed in order to determine the thermal spin wave properties ($\hbar\omega_{T}({q}),\eta_{T}({q})$). These correlations are represented in the frequency and momentum domain as the dynamical structure factor (DSF) \cite{ASD_dyna_reference}
\begin{equation}
\label{dyna}
    S(q,\Omega) = \frac{1}{N_{s}}\sum_{i,j}e^{-iqr_{ij}} \int_{-\infty}^{+\infty}dt e^{-i\Omega t} C_{ij}^{\pm}(t),
\end{equation}
where $C_{ij}^{\pm}(t) = \langle m_{i}^{+}(t)m_{j}^{-}(0)\rangle -\langle m^{+}_{i}(t)\rangle\langle m^{-}_{j}
(0)\rangle$ 
is the correlation function, $\langle \cdot\rangle$ the thermal average and $N_{s}$ the number of sites. Considering the quantization axis along $+z$, the components of the magnetic moments are combined in the following way
\begin{align}
 m_{i}^{\pm}(t)=m_{i}^{x}(t) \pm im_{i}^{y}(t),   
\end{align}
to represent transverse spin excitations. In the quasi-particle limit, where the spin waves have well-defined energies and relax exponentially to thermal equilibrium, we can identify the Lorentzian line-shape of $S(q,\Omega)$ by the fit
\vspace{0.1cm}
\begin{equation}
\label{fit_lor}
    L(\Omega;A,\omega(q),\eta({q})) = \frac{A}{\pi}\left[\frac{\eta(q)}{(\Omega-\omega({q}))^{2}+\eta({q})^{2}}\right],
\end{equation}
with the peak positions $\omega({q})$ and line-widths $\eta({q})$.

\subsection{Spin wave interactions and scattering}
  
With the introduction of temperature there will be changes to the spin wave dispersion and scattering rate due to the bath and spin wave interactions. We adapt the analytical description for $\hbar\omega_{T}(q)$ and $\eta_{T}$(q) from the work of Mclean and Blume \cite{McleanBlume_lin_chain}, who treat a linear one dimensional Heisenberg spin chain without an external magnetic field or Gilbert damping. The linear spin wave dispersion is described via the expression 
\begin{align}
\label{energy_renormalization}
    \hbar\omega_{T}(q) &= \frac{4J_{T}(1-\cos(qa)) + \mu_{m}B_{z}}{1+\alpha^{2}},
\end{align}
which depends on the temperature via $J_{T}=J_{1}\cdot\Delta_{T}$. $J_{T}$ and $\Delta_{T}$ are the Heisenberg exchange and local order parameter of the magnetization, respectively. The local order parameter $\Delta_{T}$ describes the average magnetization in some localized region (cluster) of spins. When the cluster size is extended to the entire spin chain we get that $\Delta_{T} = \braket{m(T)}$, where $\braket{m(T)}$ represents the global average magnetization. At this limit, the order parameter describes the fully ordered state when $\Delta_{T}=1$ and the fully disordered state when $\Delta_{T}=0$. The local order parameter will diminish as a function of temperature (see Fig.~\ref{local_average}) and lead to a reduced effective field for the spins in its cluster. This in turn softens the spin wave dispersion for non-zero momenta. \\
\\
The scattering rate can be separated into two contributions: renormalization of the effective damping by the bath and fluctuations in the local magnetic order. These can be represented in the following way
\begin{align}
    \label{scattering_rate}
    \eta_{T}(q)= \alpha_{T}\omega_{T}(q) + \beta_{T}\cdot\sqrt{(1-\cos(qa))}.
\end{align}
where the first term is proportional to the temperature dependent dispersion of Eq.~(\ref{energy_renormalization}), which is quadratic in momentum $q$ close to $q=0.0$. $\alpha_{T}$ is expected to be constant at low temperatures but might change for higher temperatures. The second term describes the interaction between spin waves and the local magnetic order (SW-LOI). This term  corresponds to a linear contribution in $\eta_{T}(q)$ at low $q$, proportional to $\beta_{T}$. To the best of our knowledge, the analytical expression of $\beta_{T}$ is unknown. Both $\alpha_{T}$ and $\beta_{T}$ are fitted in part from the quadratic and the linear contribution to $\eta_{T}(q)$. The extraction of these quantities will allow an estimate of the change in the bath-mediated and SW-LOI  scattering for different temperatures. 

\section{Numerical methods}
To study the effect of temperature on the spin wave dispersion and scattering rate, we need to include a temperature dependent term into the LLG. The temperature-dependent spin dynamics, as described by the stochastic Landau-Lifshitz-Gilbert (sLLG) equation, is used
\begin{equation}
\label{sllg}
    \frac{d\vec{m}_{i}}{dt} = -\frac{1}{(1+\alpha^{2})\hbar}(\vec{m}_{i} \times ((\vec{H}^{eff}_{i}+\vec{\zeta}_{i}) + \alpha(\vec{m}_{i} \times(\vec{H}^{eff}_{i}+\vec{\zeta}_{i}))),
\end{equation}
where $\vec{\zeta}_{i}$ is a stochastic field which is added to the effective field and has the following two properties:
\begin{equation}
 \label{stocastic_field}
    \langle\zeta_{i}\rangle = 0, \hspace{5mm} \langle \vec{\zeta}^{\mu}_{i}(t) \vec{\zeta}^{\nu}_{j}(t')\rangle = \frac{2\alpha k_{B}T \Delta t}{\hbar}  \delta_{ij}\delta_{\mu\nu},
\end{equation}
where $k_{B}$ is the Boltzmann constant and $\Delta t$ the numerical integration timestep. $\delta_{ij}$ and $\delta_{\mu\nu}$ are delta functions where $(i,j)$ are the site indices and ($\mu,\nu={x,y,z}$) the cartesian coordinates. We use the classical Rayleigh-Jeans statistics resulting in Eq.~(\ref{stocastic_field}) and therefore do not have a Bose Einstein distribution but rather $\bar{n}({q})=\frac{k_{B}T}{\hbar\omega({q})}$ for the spin waves \cite{Barker_quantum_thermo}. \\
\\
The ensemble average for the spin wave excitations in Eq.~(\ref{dyna}) can be replaced by a time average since $\zeta_{i}$ ensures ergodicity in the long time limit. The initial conditions for the spins are obtained from Monte Carlo simulations, for each temperature, to ensure proper equilibration. Since we are working in an one dimensional spin chain without magneto-crystalline anisotropy there is no Curie temperature. We therefore compare all the simulated temperatures to the so-called reduced temperature $T^{*}=2J_{1}/k_{B}$ \cite{Loveluck_windsor_Mori_fraction}. The $S(q,\Omega)$ is computed at each temperature, from individual time trajectories. $S(q,\Omega)$ is then averaged, over at most 400 time trajectories, to produce well-defined spectral lines. The parameters $\omega({q})$ and $\eta({q})$, in Eq.~(\ref{fit_lor}), are obtained by a fit when $S(q,\Omega)$ is converged. A description of the convergence of the DSF is detailed in Appendix \ref{sec:appendix_convergence_dyna} and the procedure to fit the parameters in Section \ref{sec:Results}. In the remainder of the paper we will refer to the averaged dynamical structure factor $\bar{S}(q,\Omega)$ which represents $S(q,\Omega)$ with the most amount of averaging unless stated otherwise.

\section{Results}
\label{sec:Results}
\begin{figure}[tbp]
    \centering
    \includegraphics[width=\linewidth]{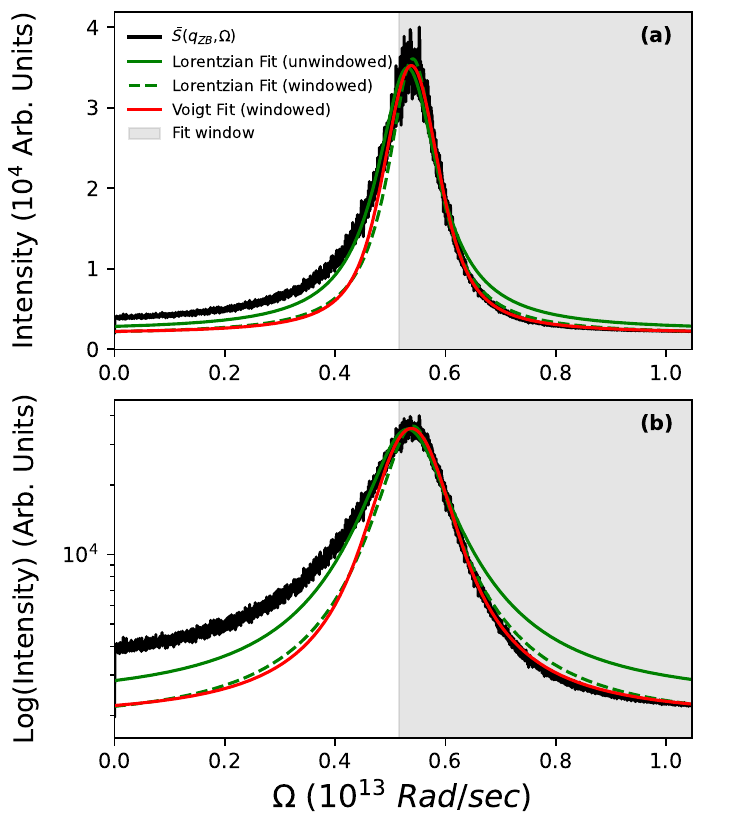}
    \caption{Averaged dynamical structure factor $\bar{S}(q_{ZB},\Omega)$, at $T=0.5 \hspace{1mm} T^{*}$ and $q_{ZB}=0.5 \ (2\pi/a)$, with different fitting approaches. (a) The Lorentzian fit is obtained from the full frequency range (solid green curve) and the windowed fit (dashed green curve).  The Voigt fit is applied in the windowed region (solid red curve). (b) Comparison of the same Lorentzian and Voigt fits on a logarithmic scale. We see that the unwindowed Lorentzian fit does not fit very well to either side of the peak. With the application of the window we see that the Lorentzian fit is improved but 
    the windowed Voigt fits the right side of the peak more accurately than the windowed Lorentzian.}
    \label{zone_edge_voigt_fit}
\end{figure}

The spin dynamics simulations are performed for a 1D chains of 64 spins ($N_{s}$) with a nearest-neighbor interaction $J_{1}$ of 0.5 meV, a magnetic field $B_{z}$ of 5 Tesla, Gilbert damping $\alpha$ of 0.05 and periodic boundary conditions. The sLLG is integrated with the atomistic spin dynamics code Matjes \cite{Matjes}, using the HeunP integrator \cite{HeunP}. An integration timestep of $\Delta t =$ 1.0 femtosecond, a sampling rate of $\Delta t_{s} =$ 150 femtoseconds and a total integration time of 3.0 nanoseconds were used for the spin dynamics. Increasing the number of sites, integration time or decreasing the sampling rate and integration timestep did not change the trends or magnitude of the obtained scattering rates and dispersions in a discernible way.\\
\\
\begin{figure}[tbp]
    \centering
    \includegraphics[width=\linewidth]{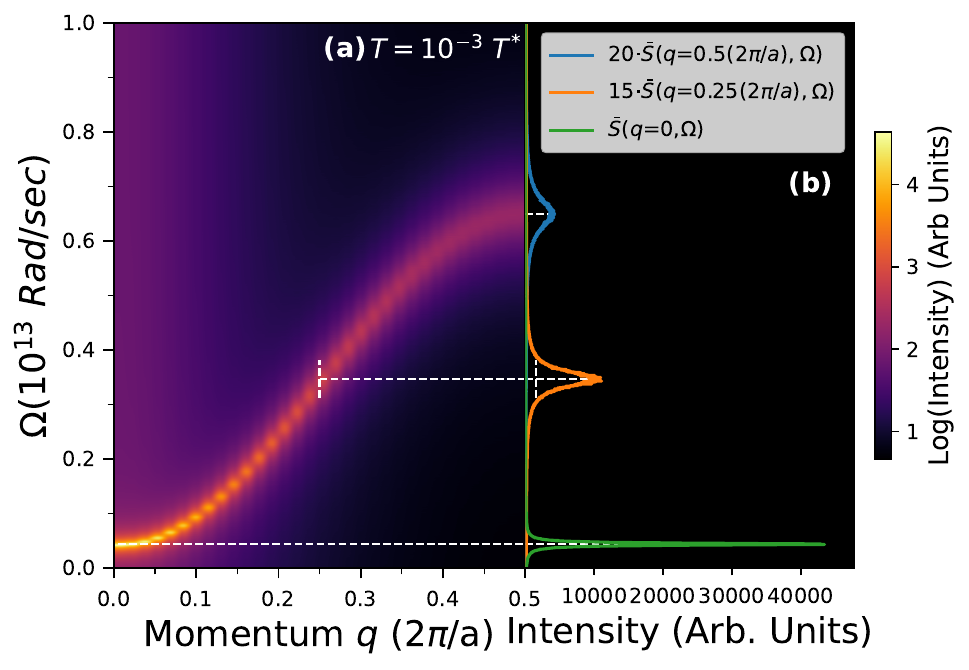}
    \caption{Averaged dynamical structure factor $\bar{S}(q,\Omega)$ at a temperature of $T = 10^{-3} \hspace{1mm} T^{*}$. (a) The dispersion increases from $q=0.0$ which corresponds to a homogeneous excitation to $q=0.5 \ (2\pi/a)$ which has a periodicity of two unit cell. The presence of a gap at $q=0.0$ is due to the external magnetic field ($B_{z}$). (b) Shows three examples of individual momentum mode $q=0.0$, $q=0.25 \ (2\pi/a)$ and $q=0.5 \ (2\pi/a)$ in green, orange and blue, respectively. Each $q$ mode can be analyzed based on the dynamical structure factor $\bar{S}(q,\Omega)$ from which the peak position $\omega({q})$ and line-width $\eta({q})$ can be determined which is represented by the horizontal and vertical dashed lines respectively.} 
    \label{averaged_dyna}
\end{figure}
Fig.~\ref{averaged_dyna} shows the averaged dynamical structure factor $\bar{S}(q,\Omega)$ as a function of momentum ($q$) in the range $[0,0.5]$, in units of $2 \pi /a$, at a very low temperature of $T=10^{-3} \hspace{1mm} T^{*}$. Fig.~\ref{averaged_dyna}(a) shows the intensity of $\bar{S}(q,\Omega)$ on a logarithmic scale. The placement of highest intensity, per momenta q, evolves as a cosine for increasing momenta and resembles the spin wave dispersion in Eq.~(\ref{dispersion_T0}) as expected. Fig.~\ref{averaged_dyna} (b) shows three examples of $\bar{S}(q,\Omega)$.  As the momentum increases, the peak intensity decreases, as observed experimentally in all materials, e.g. in neutron scattering measurements \cite{Neutron_scattering_EuS}. For increasing momentum there is an increase in the position and line-width of $\bar{S}(q,\Omega)$. This change in line-width corresponds to the scattering rate defined in Eq.~(\ref{scattering_rate_T0}).\\
\\
\begin{figure}[tbp]
    \centering
    \includegraphics[width=\linewidth]{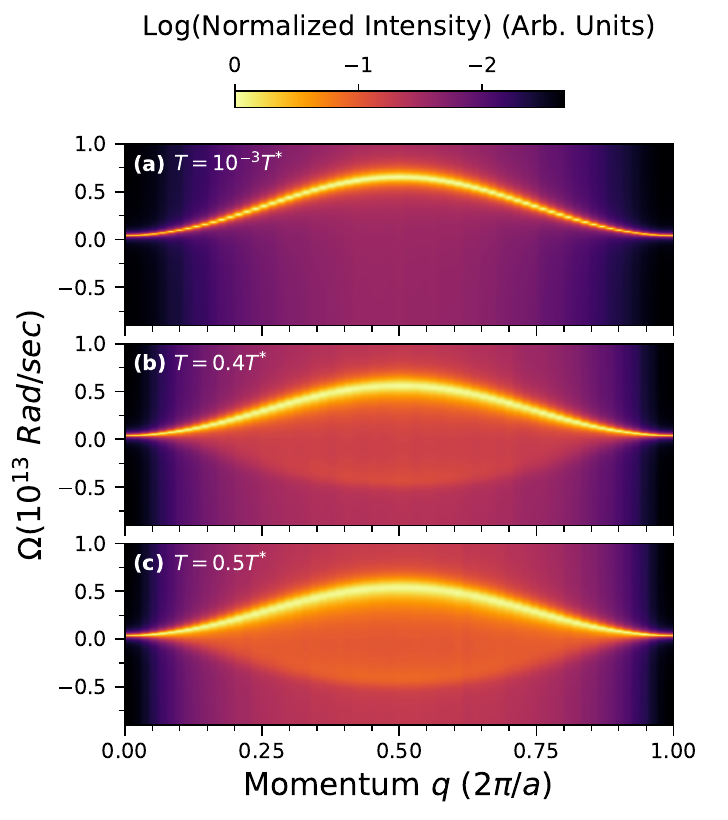}
    \caption{Averaged dynamical structure factor $\bar{S}(q,\Omega)$ for different temperatures with respect to $T^{*}$. (a) Shows the low temperature limit where the cosine dispersion and broadening appear. (b)-(c) Shows, with increasing temperature, the reduction of the dispersion, broadening of the peaks and emergence of intensity around and below zero frequency. Note that, at each $q$, the $\bar{S}(q,\Omega)$ are normalized to a maximum intensity of one, to highlight relative differences between $q$ modes.} 
    \label{Temp_dyna}
\end{figure}
For higher temperatures the $\bar{S}(q,\Omega)$ are presented in Fig.~\ref{Temp_dyna} as a function of the momentum ($q$) in the range $[0,1]$, in units of $2\pi/a$, for three different temperatures. Fig.~\ref{Temp_dyna}(a) shows $\bar{S}(q,\Omega)$ at $T = 10^{-3} \hspace{1mm} T^{*}$. The data is identical to the one presented in Fig. \ref{averaged_dyna}(a) on a logarithmic scale however the intensities of $\bar{S}(q,\Omega)$ are normalized to one, at each momentum $q= \mathrm{cte}$, to highlight relative differences between the different modes. The shape of the intensity plot highlights the frequencies being occupied and the vertical broadness indicates the line-widths. For Fig.~\ref{Temp_dyna} (b/c) we see the $\bar{S}(q,\Omega)$ at $T = 0.4 \hspace{1mm} T^{*}$ and $T = 0.5 \hspace{1mm} T^{*}$, respectively. The overall behavior of the DSF changes as temperature increases. The frequencies $\omega_{T}(q)$ decrease and scattering rates $\eta_{T}(q)$ increase for increasing temperatures. In addition to these changes we see a new branch appearing at negative frequencies. This feature does not impact our results and is further discussed in Appendix \ref{sec:negative_frequencies}.\\ 
\\
To highlight the changes in $\bar{S}(q,\Omega)$ we can take a look at the maximum of $\bar{S}(q_{ZB},\Omega)$ at $q_{ZB}=0.5 \hspace{1mm} (2\pi/a)$. We observe that the frequencies are reduced from $\omega(q_{ZB})=6.50 \cdot 10^{12} \hspace{0.5mm} Rad/sec$ to $\omega(q_{ZB})=5.38 \cdot 10^{12} \hspace{0.5mm} Rad/sec$ at $T=10^{-3} \hspace{1mm} T^{*}$ and $T =0.5 \hspace{1mm} T^{*}$ respectively. This reduction corresponds to the lowering of the effective magnetic exchange as the temperature increases and is consistent with previous work \cite{Rozsa2017-wj}. The line-width of $\bar{S}(q_{ZB},\Omega)$ increases from $\eta(q_{ZB})=3.23 \cdot 10^{11} \hspace{0.5mm} Rad/sec$ to $\eta(q_{ZB})=4.20 \cdot 10^{11} \hspace{0.5mm} Rad/sec$, at $T=10^{-3} \hspace{1mm} T^{*}$ and $T =0.5 \hspace{1mm} T^{*}$ respectively, indicating that the spin waves are scattering more for higher temperatures. \\
\\
To inspect the spectral line shape for these higher temperatures we take a closer look at $\bar{S}(q_{ZB},\Omega)$, at $q_{ZB}=0.5 \ (2\pi/a)$ and $T=0.5 \hspace{1mm} T^{*}$, which is  presented in Fig.~\ref{zone_edge_voigt_fit}(a) as the black curve. We can see that $\bar{S}(q_{ZB},\Omega)$ has a well defined peak. There is however an asymmetry in intensity around the main spin wave peak, with a faster decay at higher frequencies (right side of the peak) and a slower decay at lower frequencies (left side of the peak). This asymmetry and the presence of a diffuse regime around $\Omega \approx 0$ brings into question whether or not a single Lorentzian fit function can describe the spectral line for the full frequency range \cite{Loveluck_windsor_Mori_fraction,Paramagnetic_spin_waves_central_intenisty,Heller_blume_lin_chain,Windsor_Locke_Wheaton}. 

\begin{figure}[tbp]
    \centering
    \includegraphics[width=\linewidth]{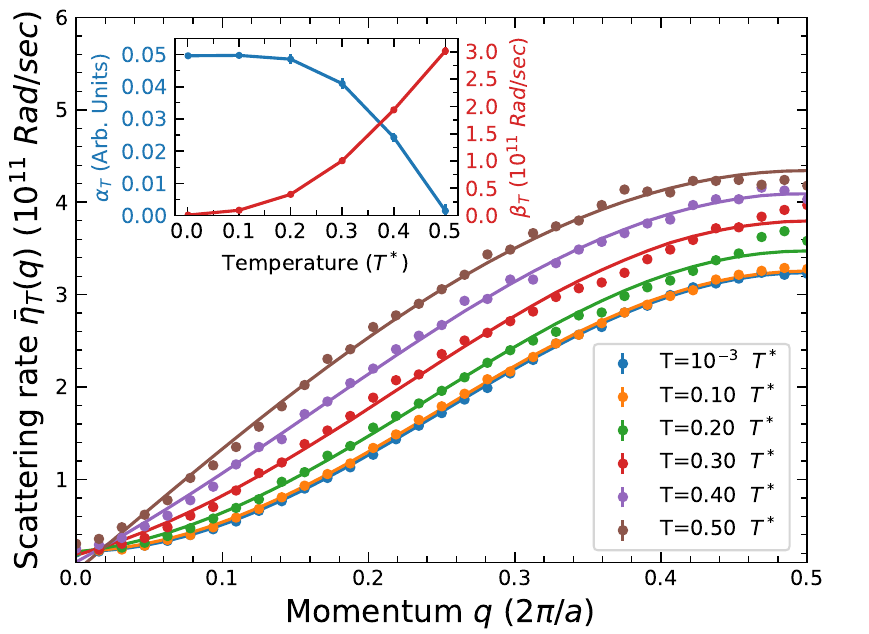}
    \caption{Temperature dependent scattering rates $\bar{\eta}_{T}(q)$ from the Voigt fits. We observe the transition from the low temperature scattering rate ($T = 10^{-3} \hspace{1mm} T^{*}$, blue curve), which coincides with Eq.~\ref{scattering_rate_T0}, towards a linearly scaling scattering rate for low momenta at the highest temperature ($T=0.5\hspace{1mm}T^{*}$, brown curve). The data points represent the average of the fitted line-widths by Eq.~\ref{avg:first}. The error-bars are within the point size. The solid lines are the fit results of $\eta_{T}(q)=\alpha_{T}\cdot\omega_{T}(q) + \beta_{T}\cdot\sqrt{1-\cos(qa)}$ to $\bar{\eta}_{T}(q)$, where $\alpha_{T}$ and $\beta_{T}$ are the fit parameters which are shown in the inset as points (solid lines in the inset are guides to the eye). The error-bars in the inset are from lmfit.}
    \label{scatter_fit}
\end{figure}

\begin{figure}[tbp]
    \centering
    \includegraphics[width=\linewidth]{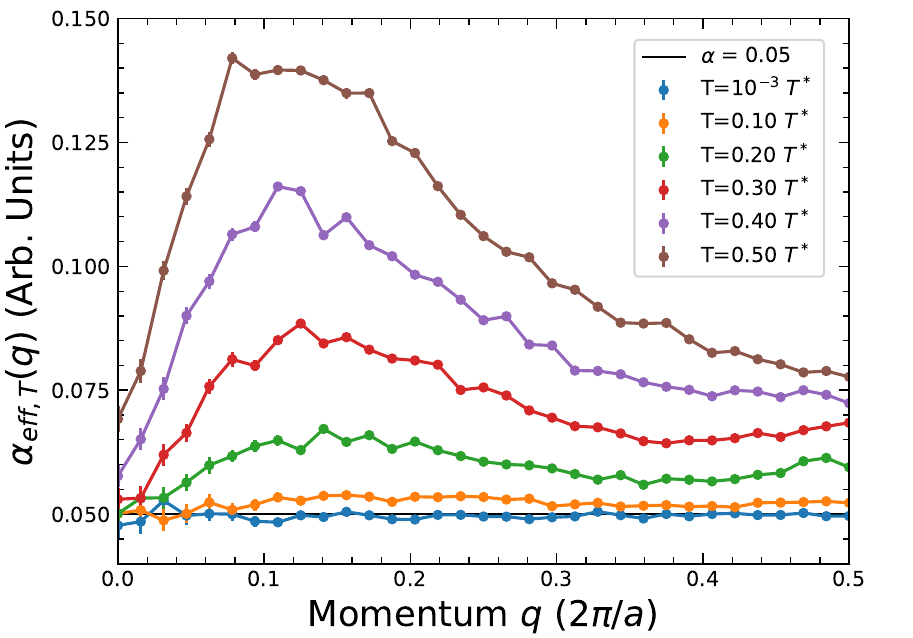}
    \caption{Temperature dependent effective Gilbert damping constructed from the scattering rates $\bar{\eta}_{T}(q)$ and  dispersions $\bar{\omega}_{T}(q)$ as $\alpha_{eff,T}(q) = \bar{\eta}_{T}(q)/\bar{\omega}_{T}(q)$. We observe the transition from the momentum independent effective Gilbert damping at the low temperature ($T = 10^{-3} \hspace{1mm} T^{*}$, blue curve), which matches the initial Gilbert damping (0.05), towards the momentum dependent effective Gilbert damping at the highest temperature ($T = 0.5 \hspace{1mm} T^{*}$, brown curve). The effective Gilbert damping increased almost threefold for certain intermediate momenta. $\alpha_{eff,T}(q)$ is represented by the data points. The solid lines are a guide to the eye.}
    \label{effective_gilbert_damping}
\end{figure}

\begin{table}[tbp] 
\begin{tabular}{|c|c|c|c|}
\hline
Fit& window&$R^{2}$& $\chi^{2}_{reduced}$\\
\hline
 $L$ & no &0.961 &  $9.95 \cdot 10^{5}$\\
 $L$& yes&0.996 & $1.92 \cdot 10^{5}$\\
 $V$ & yes&0.997&  $1.31 \cdot 10^{5}$\\
\hline
\end{tabular}
\caption{Fit statistics obtained from $\bar{S}(q_{ZB},\Omega$) at a temperature of $0.5T^{*}$ for the (un)windowed Lorentzian ($L$) and Voigt ($V$) fit.}
\label{table_Rsquare}
\end{table}
To account for the asymmetry around the excitation peak, we introduce a fit window in the following way : $\tilde\omega({q})-L\cdot \tilde\rho({q})\leq\Omega \leq \Omega_{max}$, where $\tilde\omega({q})$ and $\tilde\rho({q})$ are initial guesses from the fit library lmfit for the peak position and full width half maximum (FWHM), respectively \cite{lmfit}. The $L$ parameter is chosen in such a way that the window captures the FWHM of the peak. Note that an additional constant background offset is used in the fitting procedure. The effect of the fit window on the Lorentzian fit is shown in Fig. \ref{zone_edge_voigt_fit}(a). Without the fit window, the Lorentzian fit (solid green curve) overshoots the intensity of $\bar{S}(q_{ZB},\Omega)$ on the right side of the peak which results in the residual of $R^2=0.961$ and reduced chi-squared of $\chi_{red}^{2}=9.95 \cdot 10^{5}$. On the other hand, including the window in the Lorentzian fit (green dashed curve) allows the fit to better describe the right side of the peak resulting in the improvement of the residual $R^2=0.996$ and $\chi_{red}^{2}=1.92 \cdot 10^{5}$. The right tail however is still not fully described by the Lorentzian decay. Another fit function that could be considered here is the Voigt function. The Voigt convolves a Lorentzian together with a Gaussian profile and is defined as  
\begin{align}
    V(\Omega;\sigma({q}),\eta({q})) &\equiv \int_{-\infty}^{+\infty}G(\Omega';\sigma({q}))L(\Omega-\Omega';\eta({q}))d\Omega',\\ \nonumber
    G(\Omega;\sigma({q})) &= \frac{e^{-\Omega^{2}/2\sigma({q})^{2}}}{\sqrt{2\pi}\sigma({q})},
\end{align}
where $G(\Omega;\sigma({q}))$ is the Gaussian profile and $L(\Omega,\eta({q}))$ the Lorentzian profile defined in Eq.~(\ref{fit_lor}). The peak position offset $\omega(q)$ is left out for clarity. In Fig.~\ref{zone_edge_voigt_fit}(b) we see the Lorentzian and Voigt fits on a logarithmic scale with the fit window. The Voigt fit (solid red curve) shows an additional improvement in fit for the right tail, with respect to the Lorentzian fit, with a residual of $R^{2}=0.997$ and $\chi_{red}^{2}=1.31 \cdot 10^{5}$. We can look at all the fit statistics together, reported by lmfit, to determine which fit function performed the best beyond visual inspection. These fit statistics are summarized in Table \ref{table_Rsquare} from which we can conclude that the Voigt fit describes the data of $\bar{S}(q_{ZB},\Omega)$, at $T = 0.5 \hspace{1mm} T^{*}$, the best since it has the highest $R^{2}$ and lowest $\chi^{2}_{reduced}$. An overview of the observables from the Lorentzian and Voigt fits are presented in Table \ref{table_Voigt_lorentzian_fit_results}, where
\begin{table}[tbp]
\begin{tabular}{|c|c|c|c|c|c|}
\hline
Fit& window&$\omega(q_{ZB})$ & $\eta(q_{ZB})$ & $\sigma(q_{ZB})$ & FWHM($q_{ZB}$)\\
\hline
 $L$ & no &$5.31 \cdot 10^{12}$ & $6.79 \cdot 10^{11}$ & 0 & $1.36 \cdot 10^{12}$\\
 $L$& yes&$5.41\cdot 10^{12}$ & $5.48\cdot 10^{11}$  & 0&$1.10\cdot10^{12} $\\
 $V$& yes&$5.38\cdot10^{12}$& $4.20\cdot10^{11}$ & $2.84\cdot10^{11}$ & $1.22\cdot10^{12}$ \\
\hline
\end{tabular}
\caption{Fit parameters obtained from $\bar{S}(q_{ZB},\Omega$) at a temperature of $0.5T^{*}$ for the (un)windowed Lorentzian ($L$) and Voigt ($V$) fit.}
\label{table_Voigt_lorentzian_fit_results}
\end{table}
we see that from the unwindowed to the windowed Lorentzian the peak position $\omega(q_{ZB})$ increases. From the windowed Lorentzian to windowed Voigt the peak position $\omega(q_{ZB})$ slightly decreases. The changes in peak positions are relatively minor when compared to the changes in line-width. The inclusion of the window for the Lorentzian fit reduces $\eta(q_{ZB})$ by about 19.3\%. The change from a windowed Lorentzian to a windowed Voigt further reduces $\eta(q_{ZB})$ by about 23.4\%. This line-width difference between the Lorentzian and Voigt fit is due to the allowed redistribution of the FWHM between $\eta(q_{ZB})$ and $\sigma(q_{ZB})$ for the Voigt with respect to the Lorentzian. The necessity of the Voigt function to describe the spectral line shape could originate from the so called inhomogeneous line-broadening. This is where the Voigt function arises from the sum of normally distributed Lorentzian line shapes. Other explanations might be the interaction with a wide variety of spin wave excitations, or with the external thermal bath. However, the study of the origin of this Voigt function is beyond the scope of this work. To obtain the temperature-dependent dispersions and scattering rates, we systematically use a Voigt fit, where the reported line-width ($\eta_{T}(q)$) are the Lorentzian contribution, which should correspond to the intrinsic inverse lifetime of the quasi-particles. \\
\begin{figure}[tbp]
    \centering
    \includegraphics[width=\linewidth]{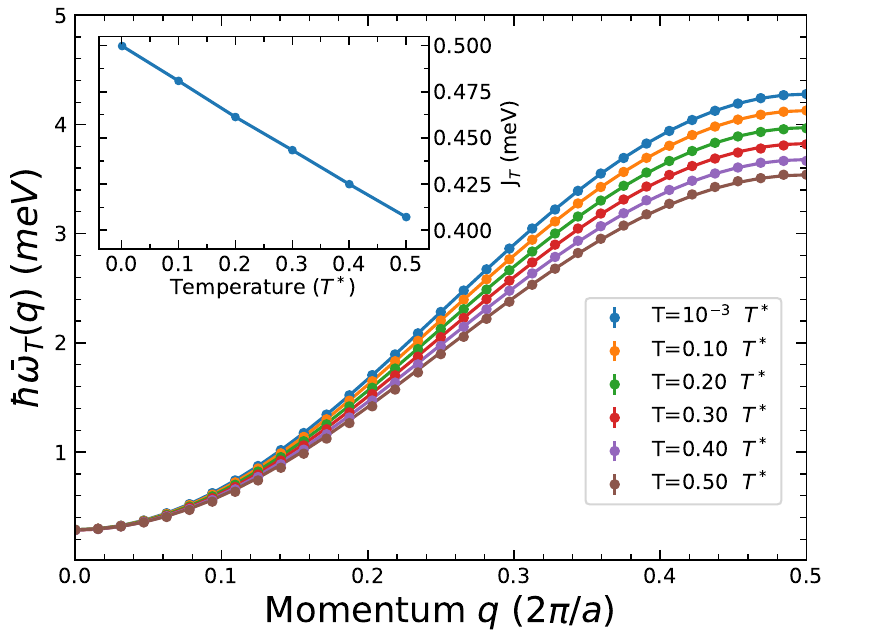}
    \caption{Temperature dependent dispersions $\hbar\bar{\omega}_{T}(q)$ obtained from the Voigt fits. We observe the expected dispersion for $T = 10^{-3} \hspace{1mm}T^{*}$ (blue curve) which coincides with Eq.~\ref{dispersion_T0}. For higher temperatures we see the progressive softening of the dispersion. The data points represent the averaged fitted peak positions by Eq.~\ref{avg:first}. The error-bars are within the point size. The solid lines are the fit results of $\hbar\omega_{T}(q)=(1+\alpha^{2})^{-1}\cdot[4J_{T}\cdot(1-\cos(qa)) + \mu_{m}B_{z}]$ to $\hbar\bar{\omega}_{T}(q)$, where $J_{T}$ is the fit parameter which is shown in the inset as points (solid lines are a guide to the eye). The error-bars in the inset, from lmfit, are within the point size.}
    \label{linear_frequency_fit}
\end{figure}
\\
The temperature-dependent spin wave dispersions $\hbar\bar{\omega}_{T}(q)$, as a function of momentum ($q$) and temperature ($T$), are shown in Fig.~\ref{linear_frequency_fit}. The temperature is represented in terms of the reduced temperature $T^{*} = 2J_{1}/k_{B}$. The points represent the peak positions $\hbar\bar{\omega}_{T}(q)$, as outlined by Eq.~(\ref{avg:first}), and are obtained from the fits of the Voigt function to $\bar{S}(q,\Omega)$ at different temperatures. The definition of the error-bar is in Appendix \ref{sec:appendix_convergence_dyna}. The solid lines correspond to the fits of Eq.~(\ref{energy_renormalization}) to $\hbar\bar{\omega}_{T}(q)$. At the lowest temperature $T= 10^{-3} \hspace{1mm} T^{*}$ (blue curve) we observe the dispersion increasing, as a function of increasing $q$, with the minimum at $q=0.0$ (external magnetic field energy) and maximum at $q=0.5 \ (2\pi/a)$. For a higher temperature at $T = 0.1 \hspace{1mm} T^{*}$ (orange curve) we see the dispersion being lower, for all momenta except $q=0.0$, with respect to the dispersion at $T = 10^{-3} \hspace{1mm} T^{*}$. For increasingly higher temperatures the dispersion keeps lowering without changing its momentum dependence. To see this preservation of the momentum dependence we need to analyze the $J_{T}$ parameters from the fits of Eq.~(\ref{energy_renormalization}) to the dispersions. These parameters are shown as points in the inset of Fig. \ref{linear_frequency_fit}. The effective exchange $J_{T}$, which is a prefactor of the $q$ dependent term in Eq.~(\ref{energy_renormalization}), starts approximately at the initial exchange value $J_{1}$ (0.5 meV) for the lowest temperature $T = 10^{-3} \hspace{1mm} T^{*}$. For increasingly higher temperatures we see that $J_{T}$ keeps reducing down to an eventual value of 0.407 meV at $T = 0.5 \hspace{1mm} T^{*}$. The reduction of $J_{T}$, which is determined by the change in the local order parameter $\Delta_{T}$ (since $J_{T}=J_{1}\cdot \Delta_{T}$), captures the observed lowering of the dispersions without altering the momentum dependence of the dispersion. To verify the temperature dependence of $\Delta_{T}$ itself we compute the spatially and temporally averaged magnetization $|M(N_{c},T)|$ where $N_{c}$ refers to the number of spins, within a cluster, included in the spatial average.
\begin{figure}[tbp]
    \centering
    \includegraphics[width=\linewidth]{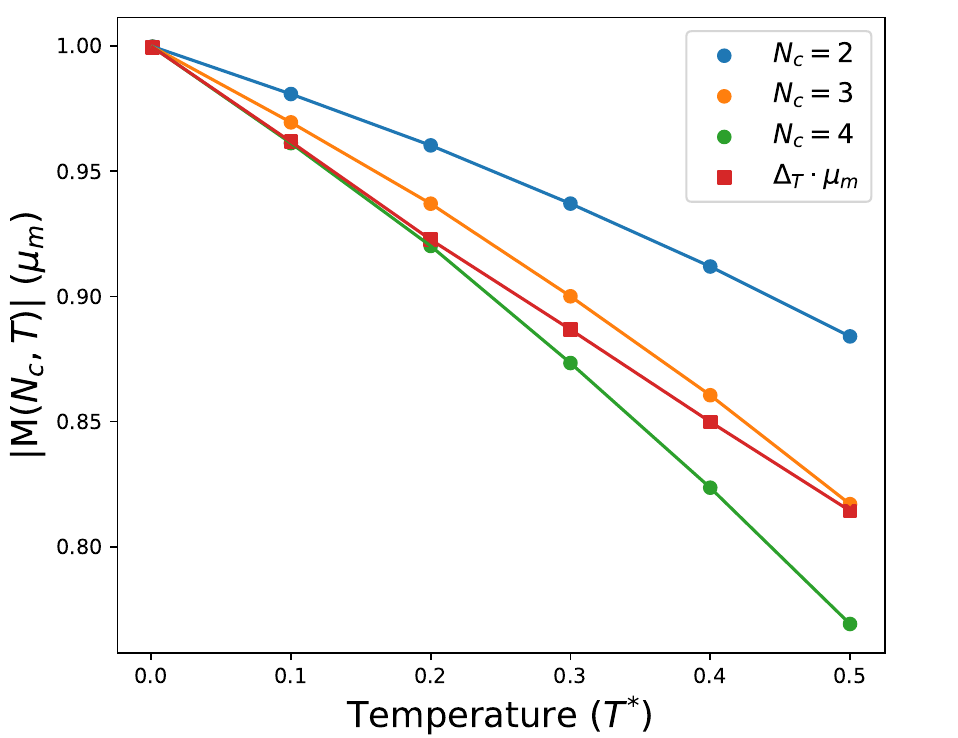}
    \caption{Averaged local magnetization $|M(N_{c},T)|$ for different reduced temperatures $T^{*}$ and cluster size $N_{c}$. The fitted $\Delta_{T}$ lies in between the three ($N_{c}$ = 3) and  four ($N_{c}$ = 4) spin clusters except for the lowest temperature where it is below both curves. For increasing temperatures we observe a transition in interaction length from next nearest neighbor ($N_{c}= 4$) to nearest neighbor ($N_{c} = 3$).}
    \label{local_average}
\end{figure}
Fig.~\ref{local_average} shows the averaged magnetization $|M(N_{c},T)|$ as points for different temperatures and cluster sizes $N_{c}$ (lines are a guide to the eye). For $N_{c}=2$ (blue curve) we see the decrease of the averaged local magnetization from $|M(N_{c},10^{-3} \hspace{1mm} T^{*})|$ $\approx 1$ to $|M(N_{c},0.5 \hspace{1mm} T^{*})|$ $\approx 0.88$. With the inclusion of more spins in the cluster ($N_{c}=3,4$, orange and green curve respectively) we see that the decrease of $|M(N_{c},T)|$, with increasing temperatures, becomes larger for bigger spin clusters. The local order parameter fits $\Delta_{T}$ (red curve) lie in between the three and four spin cluster average (orange and green curve) for all temperatures except $T = 10^{-3} \hspace{1mm} T^{*}$ where it lies below both curves. We can see that $\Delta_{T}$ transitions from a scaling similar to the 4 spin cluster for $T = 0.1,0.2 \hspace{1mm} T^{*}$ towards a 3 spin cluster scaling at $T = 0.5 \hspace{1mm} T^{*}$. This shows that the representative cluster size decreases as the temperature increases, and that around one to two interaction shells (three to four spins) are necessary to accurately describe the local order parameter, and in turn the effective field, for most of the temperature range. We conclude that the softening of the spin wave dispersions is caused by the reduction of the local order parameter, which scales as the averaged local magnetization between the three and four spin cluster. Spin wave dispersion softening with temperature has been observed many times in the literature \cite{McleanBlume_lin_chain,Windsor_Locke_Wheaton,Magnon_softening_DFT,Barker_YIG_dyna,Rozsa2017-wj}. \\
\\
Fig. \ref{scatter_fit} shows the scattering rates  $\bar{\eta}_T(q)$ as a function of momentum ($q$) and temperature ($T$). The points corresponds to the scattering rates  $\bar{\eta}_T(q)$, as outlined by Eq.~(\ref{avg:first}), and are obtained from the fits of the Voigt function to $\bar{S}(q,\Omega)$ at different temperatures. The lines correspond to fits of Eq.~(\ref{scattering_rate}) to the scattering rates $\bar{\eta}_T(q)$. At the low temperature of $T$ = $10^{-3} \hspace{1mm} T^{*}$ (blue curve), we observe that the scattering rates increases as a function of increasing $q$. This suggests that SW will be more scattered as $q$ increases and therefore the lifetime should decrease as a function of $q$. When the temperature increases from $T$ = $10^{-3} \hspace{1mm} T^{*}$ to $T$ = $0.1 \hspace{1mm} T^{*}$ (orange), the trend of the scattering rates barely changes and both curves lie on top of each other. When the temperature is up to $T$ = $0.2 \hspace{1mm} T^{*}$ (green), the scattering rates starts to increase faster than the two previous cases. This shows that the SW are more likely to scatter as the temperature increases. This trends continues and further amplifies as the temperature increases to $T$ = $0.3 \hspace{1mm} T^{*}$ (red), $T$ = $0.4 \hspace{1mm} T^{*}$ (purple) and $T$ = $0.5 \hspace{1mm} T^{*}$ (brown). Let's now turn our attention to the region of low momenta $q \leq 0.1 \ (2\pi/a)$. At the low temperature $T$ = $10^{-3} \hspace{1mm} T^{*}$, the scattering rate  $\bar{\eta}_T(q)$ seems to increase as $q^2$ around $q\approx0.0$ which is in agreement with the zero temperature scattering rate of Eq.~(\ref{scattering_rate_T0}). As the temperature increases, this trend seems to evolve from quadratic to linear at $T$ = $0.5 \hspace{1mm} T^{*}$ at low momenta.
This transition from quadratic to linear corresponds to a change in the temperature dependencies of the $\alpha_{T}$ and $\beta_{T}$ parameters in Eq.~(\ref{scattering_rate}). When $\bar{\eta}_T(q)$ has a quadratic increase close to $q=0.0$, the $\alpha_{T}$ is unchanged as compared to the initial Gilbert damping coefficient $\alpha$ present in Eq.~(\ref{sllg}). As the temperature increases the quadratic trend becomes linear which means that the second term $\beta_T$ becomes predominant due to the increased scattering of spin waves with the local order (SW-LOI). In the high temperature case, i.e. $T$ = $0.5 \hspace{1mm} T^{*}$, the quadratic contribution apparently vanishes and the scattering rate becomes mainly linear. To quantify this effect we show the coefficients $\alpha_T$ and $\beta_T$ from Eq.~(\ref{scattering_rate}), obtained by fitting $\bar{\eta}_T(q)$ for each temperature, in the inset of Fig.~\ref{scatter_fit} as blue and red points respectively. The lines are a guide to the eye. At the lowest temperature we observe that $\alpha_{T}$ is approximately equal to the Gilbert damping ($\alpha$ = 0.05) and $\beta_{T}$ is nearly zero. As the temperature increases, $\alpha_T$ stays almost constant up to $T$ = $0.2 \hspace{1mm} T^{*}$ and then drops down to zero. The fact that $\alpha_{T}$ remained constant up to $T = 0.2 \hspace{1mm} T^{*}$ implies that the frequency reduction observed in Fig. \ref{linear_frequency_fit} corresponds to a reduction in the bath contribution to the scattering rate. For higher temperatures ($T \geq 0.3 \hspace{1mm} T^{*}$) the reduction in the bath scattering is even stronger than the frequency softening, and $\alpha_T$ goes to 0. Whether or not this strong decay of $\alpha_{T}$ is physically driven or an inability to fit a cosine shape to the dominantly linear q dependence is not investigated further in this work. On the contrary to the $\alpha_{T}$ scaling, we see that $\beta_T$ increases continuously as the temperature increases. This increase of $\beta_{T}$ implies a transition towards SW-LOI being the dominant contribution to scattering. We conclude that the overall scattering rate $\bar{\eta}_{T}(q)$ increases as a function of temperature, despite the reduction in bath scattering, due to the increase in spin wave scattering with the local magnetic order and therefore transitions from a quadratic to linear momentum scaling at low momenta. \\
\\
We can recast the temperature dependencies of the dispersion and scattering rate into the effective Gilbert damping, defined as 
\begin{align}
    \alpha_{eff,T}(q) = \frac{\bar{\eta}_{T}(q)}{\bar{\omega}_{T}(q)},
\end{align}
which is plotted in Fig.~\ref{effective_gilbert_damping} as a function of momentum ($q$) and temperature ($T$). For the lowest temperature $T = 10^{-3} \hspace{1mm} T^{*}$ (blue curve) we observe that the effective damping matches approximately the momentum-independent input Gilbert damping $\alpha =0.05$. For a higher temperature at $T = 0.1 \hspace{1mm} T^{*}$ (orange curve) we see that the effective damping increased, with respect to $T = 10^{-3} \hspace{1mm} T^{*}$, for almost all momenta but no clear momentum dependence is discernible. For $T = 0.2 \hspace{1mm} T^{*}$ (green curve) we see another increase in effective damping with respect to $T = 0.1 \hspace{1mm} T^{*}$. At this temperature however we also observe that there emerges a clear momentum dependence for the effective damping. From $q=0.0$ the effective damping increases towards a maximum around $q \approx 0.14 \ (2\pi/a)$ and then decreases again towards the zone boundary at $q \approx 0.5 \ (2\pi/a)$. For the temperatures $T = 0.3 \hspace{1mm} T^{*}$ (red curve), $T = 0.4 \hspace{1mm} T^{*}$ (purple curve) and $T = 0.5 \hspace{1mm} T^{*}$ (brown curve) we see that the effective damping increases for all momenta as a function of increasing temperature. The maximum effective damping  at $T = 0.5 \hspace{1mm} T^{*}$ reaches a value of approximately 0.142 around $q \approx 0.08 \ (2\pi/a)$. The maxima in the effective damping seems to be moving from $q \approx 0.14 \ (2\pi/a)$ at $T = 0.2 \hspace{1mm} T^{*}$ towards $q \approx 0.08 \ (2\pi/a)$ at $T = 0.5 \hspace{1mm} T^{*}$. \\
\\
We conclude that the effective Gilbert damping acquires a momentum and temperature dependence resulting from spin wave scattering with the local magnetic order and a reduction in bath scattering, partially due to the softening of the spin wave dispersion. For some of the intermediate SW momenta the effective Gilbert damping can increase almost threefold, for the highest temperature, with respect to the initial Gilbert damping.

\section{Discussion}
The definition of the quasi-particle scattering rate is derived for the small-angle approximation of the spin deflection from the quantization axis (z-direction). In the higher temperature regime this approximation is not valid anymore, and we switch from a Lorentzian to a Voigt function fit, which is often used in experiments as well. The fit accuracy increases thanks to the extra degree of freedom of the convolved Gaussian width, but raises a question of interpretation for the observed FWHM from $S(q,\Omega)$. The Lorentzian fit will reproduce the FWHM to a certain extent, while the Voigt function improves upon this fit by distributing the FWHM over the Gaussian and Lorentzian widths. The Lorentzian line-width in the Voigt function corresponds to the expected exponential real time decay. The Gaussian line-width however would correspond to a real time Gaussian decay (quadratic rather than linear in time) which is not typically identified as a quasi-particle lifetime. The Lorentzian line-width extracted from the total Voigt width might not accurately describe the full scattering rate of the spin waves, but serves as a lower bound with respect to the result from the pure Lorentzian fit. The inclusion of a fit window, to circumvent the asymmetry of certain spectral line-shapes, assumes that the main excitation peak can still be described by a single Lorentzian or Voigt. The spectral intensity plateau at low (positive and negative) frequency has been seen numerically in other works, but we have not found a simple analytical form to fit for its frequency dependence.

The treatment of spin dynamics in one dimension produces some unique features in the effective spin wave damping. The extension of these damping results to higher dimensions and more realistic systems will be investigated in the future. The effective damping might give very different results, due to these extensions, since there are more scattering channels for the spin waves and the potential for a well defined global spin order. The Markovian approximation collects all external interactions (phonons, electrons, impurities etc...) into a single constant Gilbert damping in the LLG and our results may also change when considering coupled equations of motion like spin-lattice dynamics.
\clearpage

\section{Conclusion}
In this work we have investigated the applicability of the simple scattering rate model $\eta({q}) = \alpha\omega({q})$ to the one dimensional Heisenberg spin chain, with an external magnetic field and temperature propagated by the sLLG. From the dynamical structure factor we obtain the temperature-dependent dispersions and scattering rates. The spin wave dispersion softens with increasing temperature. For the scattering rate we have identified that it is no longer solely proportional to the spin wave dispersion but rather a combination of the Gilbert-damping and a scattering term. These terms have opposite temperature dependencies: the Gilbert-damping term decreases with temperature, partially due to the spin wave dispersion softening; the scattering term increases due to the presence of clusters with different local magnetic order. These temperature-dependent properties combine to produce an effective Gilbert damping: $\alpha_{eff}$ which shows a transition from a momentum-independent damping for low temperatures, to an enhanced and strongly momentum-dependent damping for higher temperatures. These effects may be seen experimentally and would have strong consequences for the optimization of low dimensional spin wave devices.
 
\begin{acknowledgments}
We thank Artim Bassant, Aloïs Castellano and Sebastian Meyer for fruitful discussions about the calculations.
The authors acknowledge the Fonds de la Recherche Scientifique (FRS-FNRS Belgium) and Fonds Wetenschappelijk Onderzoek (FWO Belgium) for EOS project CONNECT (G.A. 40007563). BD acknowledges the Fonds de la Recherche Scientifique (FRS-FNRS), B-1000 Bruxelles, Belgium.
MJV acknowledges funding by the Dutch Gravitation program
“Materials for the Quantum Age” (QuMat, reg number 024.005.006), financed by the Dutch Ministry of Education, Culture and Science (OCW).

Simulation time was provided by 
the Consortium d'Equipements de Calcul Intensif (FRS-FNRS Belgium Grant No. 2.5020.11).
\end{acknowledgments}

\appendix

\section{Convergence of the dynamical structure factor}
\label{sec:appendix_convergence_dyna}
In order to obtain converged observables we need to average the dynamical structure factor $S(q,\Omega)$ over many spin dynamics runs, initialized from different thermalized spin configurations. Initial averaging is performed over N=200 runs, then further batches of 20 runs are saved to disk and added to the average, up to $M=10$ batches. 
The averaged dynamical structure factor is defined as
\begin{align}
    \bar{S}_{M}(q,\Omega) &= \frac{1}{200+20M} \sum_{j=1}^{200 + 20M} S_{j}(q,\Omega),
\end{align}
where $S_{j}(q,\Omega)$ are from individual spin dynamics runs. For each number of batches ($M$) we fit $\bar{S}_{M}(q,\Omega)$ and obtain the observables $\omega_{M}(q)$ and $\eta_{M}(q)$, using the python module lmfit. We compute the mean and variance for the observables over all batches as
\begin{align}
    \bar{O}({q}) &= \frac{1}{10}\sum_{M=1}^{10} O_{M}(q)\label{avg:first},\\
    Var(O({q}))&= \frac{1}{10}\sum_{M=1}^{10} |({O}_{M}(q)-{\bar{O}}({q})|^{2} \label{variance:first}, 
\end{align}
where ${O}({q})$ is either $\omega(q)$ or $\eta(q)$. The estimation of the error for $O(q)$ is the standard deviation $\sigma_{STD}(O(q)) = \sqrt{Var(O({q}))}$. When the standard deviation falls below the minimal resolution given by the time Fourier transform ($\sigma_{res}=1.047\cdot 10^{9} \ Rad/sec$) then we select the $\sigma_{res}$ as the standard deviation, i.e. $\sigma_{STD}(O(q)) = max(\sqrt{Var(O({q}))},\sigma_{res})$. In Fig.~\ref{linear_frequency_fit},\ref{scatter_fit} and \ref{effective_gilbert_damping} the points represent $\bar{O}(q)$ and the error-bars $\sigma_{STD}(O(q))$. 

\section{Negative frequency intensity in $S(q,\Omega)$}
\label{sec:negative_frequencies}
\begin{figure}[b] 
    \centering
    \includegraphics[width=\linewidth]{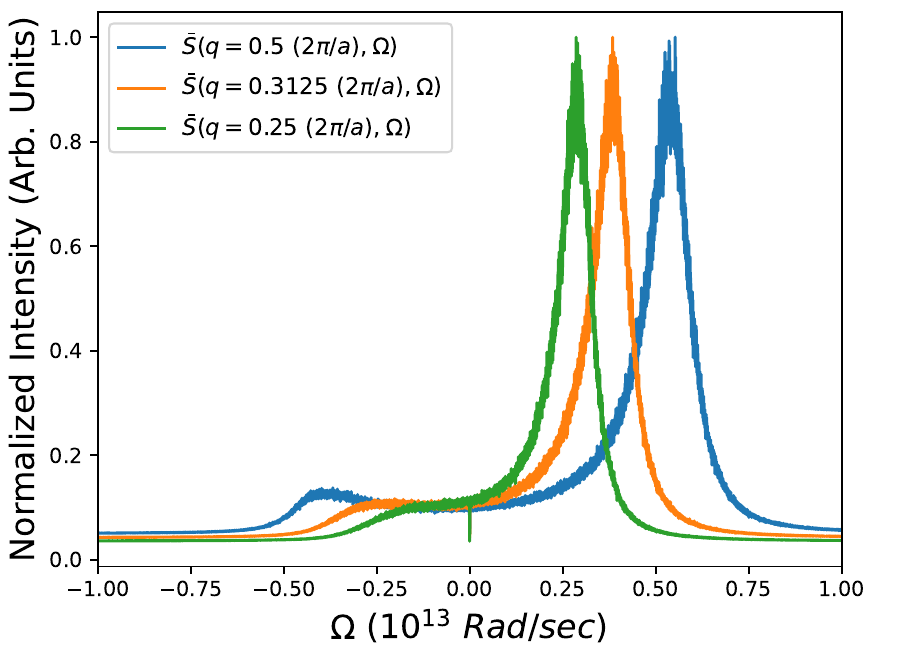}
    \caption{Negative frequency intensity in the averaged dynamical structure factor $\bar{S}(q,\Omega)$ at a temperature of $T=0.5 \hspace{1mm}T^{*}$. We see the emergence of an intensity peak for increasing momenta, from $q=0.25 \ (2\pi/a)$ (green curve) to $q=0.5 \ (2\pi/a)$ (blue curve), in the negative frequency range. The individual $\bar{S}(q,\Omega)$ intensities are normalized to unity to highlight the relative height of the main excitation peak w.r.t. the negative frequency peak.}
    \label{neg_freq_modes}
\end{figure}
At temperatures which are low w.r.t. $T^{*}$ we observe that $\bar{S}(q,\Omega)$ displays the expected dispersion and Lorentzian decay as seen in Fig.~\ref{Temp_dyna}(a). 
\begin{figure}[htp]
    \centering
    \includegraphics[width=\linewidth]{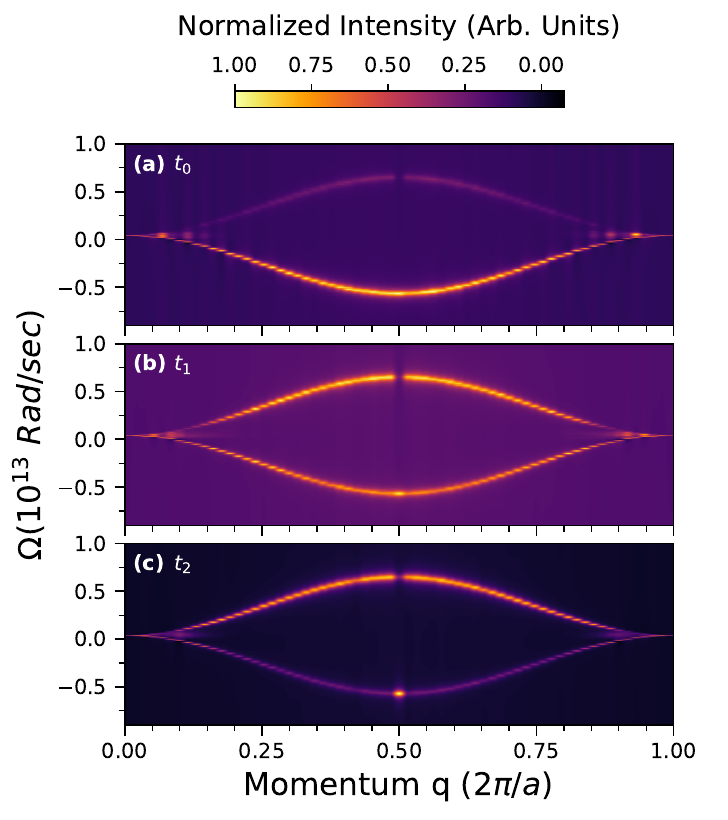}
    \caption{Averaged dynamical structure factor at a temperature of $10^{-3} \hspace{1mm} T^{*}$ from the initialized inverted SW ($q_{ZB}$) for varying simulation times. (a). Shows the prominent intensity for the negative frequencies with a simulation time $t_{0}$. (b) Highlights the roughly equal relative 
    intensity (per q mode) between the negative and positive frequencies at simulation time $t_{1}$. (c) The longest simulation time ($t_{2}$) gives a similar $\bar{S}(q,\Omega)$ to the one obtained in Fig.~\ref{Temp_dyna}(a) with a remnant occupation at negative frequencies. Note that at each q, the $\bar{S}(q,\Omega)$ are normalized to a maximum intensity of one, to highlight relative differences between q modes and different simulation lengths.}
    \label{low_temperature_inverted_state}
\end{figure}
However for higher temperatures we see the emergence of intensity for $\bar{S}(q,\Omega)$ around and below $\Omega = 0$, as seen in Fig.~\ref{Temp_dyna}(b/c). The presence of this intensity has been observed in literature before for $\Omega \geq 0$ but was not, to our knowledge, represented for the negative frequency domain (e.g. in Ref.~\cite{3D_criticality_heisenberg_chain}). We can even identify a local intensity maximum in the negative frequency domain for certain momenta of $\bar{S}(q,\Omega)$ as seen in Fig.~\ref{neg_freq_modes}. We emphasize that these are negative and not imaginary frequencies excitations. They do not correspond to instabilities of the initial ground state, but rather to short lived coherent excitations which precess counter to the ``normal'' modes. Fig.~\ref{neg_freq_modes} shows $\bar{S}(q,\Omega)$ for a few select momenta at a temperature of $T=0.5 \hspace{1mm} T^{*}$. For all three curves we can clearly identify an excitation peak in the positive frequency range. We see that for $q=0.25 \ (2\pi/a)$ (green curve) there is no identifiable peak for the negative frequency range. For $q=0.3125 \ (2\pi/a)$ (orange curve) we start to faintly see the emergence of an excitation peak 
\begin{figure}[htp]
    \centering
    \includegraphics[width=\linewidth]{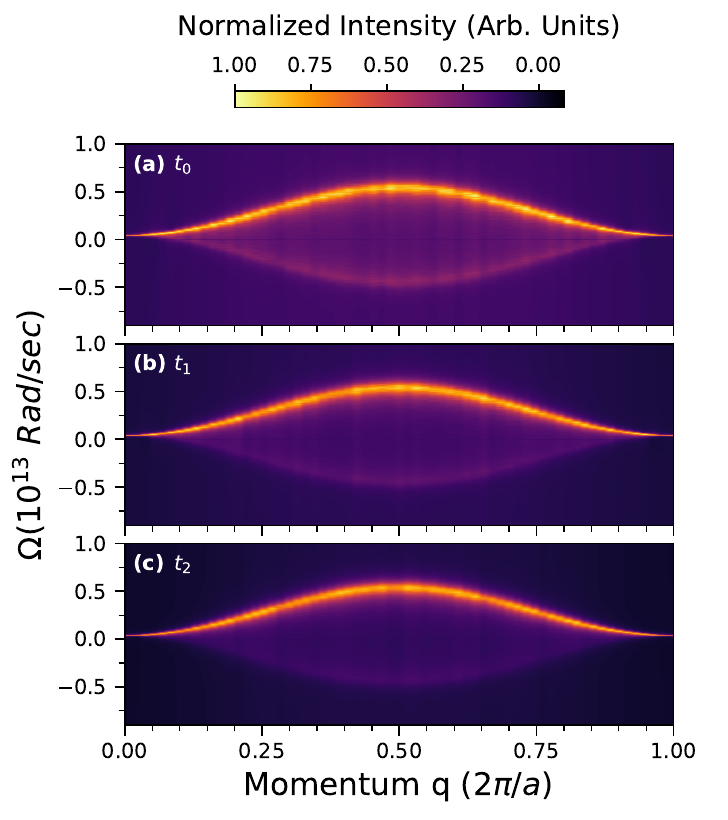}
    \caption{Averaged dynamical structure factor at a temperature of $0.5 \hspace{1mm} T^{*}$ from the initialized inverted SW ($q_{ZB}$) for varying simulation times. (a-c) Shows the change in relative intensity (per $q$ mode) from the negative to the positive frequencies. The $\bar{S}(q,\Omega)$ closely resembles the results seen in Fig.~\ref{Temp_dyna}(b/c). Note that at each q, the $S(q,\Omega)$ are normalized to a maximum intensity of one, to highlight relative differences between q modes and different simulation lengths.}
    \label{inverted_state_high_T}
\end{figure}
and for $q=0.5 \ (2\pi/a)$ (blue curve) we clearly see an excitation peak in the negative frequency range. For higher temperatures there is enough thermal energy to allow the spins to flip (to the -z direction) and occupy these unstable negative frequency modes, which precess around the opposite quantization axis (-z direction). To elucidate the origin of these modes we consider the following simulation setup. We initialize the ASD from a spin wave excitation with $q_{ZB}=0.5 \ (2\pi/a)$ in the negative z direction ($m_{i}^{z} \approx$ -1). The ASD is then propagated for two different temperatures ($10^{-3}$ $T^{*}$, 0.5 $T^{*}$) and three different integration times: $t_{0}=0.3$ ns, $t_{1}=0.6$ ns and $t_{2}=1.2$ ns. The idea is that the negative frequency modes are (un)likely to occur and keep propagating at a high (low) temperature. The different integration times will enable us to observe the return of the negative frequency modes towards thermal equilibrium and thus the shift in intensity between the negative and positive frequency peaks in $\bar{S}(q,\Omega)$ over time. $\bar{S}(q,\Omega)$ is averaged over 200 spin dynamics runs. \\

In Fig.~\ref{low_temperature_inverted_state}(a) we see the averaged dynamical structure factor $\bar{S}(q,\Omega)$ for a simulation time $t_{0}$ at a temperature of $T= 10^{-3} \hspace{1mm} T^{*}$. This integration time is not quite large enough to fully reverse the average magnetization from the negative to positive z direction and be approximately at thermal equilibrium. The intensity at negative frequencies emerges from the initialization of the inverted SW, which can decay into other negative frequency modes. The gradual return of the average magnetization to its reference state (+z) means that positive frequency modes will be populated more than the short-lived negative frequency ones at low temperature over time, which transfers relative intensity towards the positive frequencies. This shift in relative intensity can be seen in Fig.~\ref{low_temperature_inverted_state}(b) which shows the relative intensity (per $q$ mode) being roughly equal between the negative and positive frequency dispersion due to a longer simulation time ($t_{1}$). In Fig.~\ref{low_temperature_inverted_state}(c) we see the relative intensity being higher for the positive frequency dispersion w.r.t. the negative frequency dispersion due to the even longer simulation time ($t_{2}$). The lack of visible relative intensity for the positive frequency at $q_{ZB}=0.5 \ (2\pi/a)$ is due to the high initial intensity of the negative frequency at the same momenta. \\ 
\\
For the high temperature simulation ($T = 0.5 \hspace{1mm} T^{*}$) we obtain a similar transition from high to low negative frequency intensity as a function of increasing time. The final relative intensities in Fig. \ref{inverted_state_high_T}(c) is functionally the same result as for the dynamical structure factor shown in Fig. \ref{Temp_dyna}(c). These inverted spins with negative frequencies or so called anti-magnons are still an active topic of research in the current literature and may provide useful applications in quantum entanglement and spin wave amplification \cite{anti_magnons_Artim,Romling:2026ciw,Inverted_state_measurement}.

\clearpage

\bibliographystyle{unsrt}
\bibliography{references}

\end{document}